# A comprehensive and reliable protocol for manual segmentation of the human claustrum using high-resolution MRI

Steven Seung-Suk Kang[a*], Joseph Bodenheimer[b], Kayley Morris[c,] Tracy Butler[d]

[a]Department of Biomedical Sciences, University of Missouri-Kansas City, Kansas City, MO, USA
[b]Department of Psychology, University of Kansas, Lawrence, KS, USA
[c]Department of Psychology, Penn State University, University Park, PA, USA
[d]Department of Radiology, Weill Cornell Medicine, New York, NY, USA
*kangseung@umkc.edu

**Abstract**

The claustrum is a thin gray matter structure in each brain hemisphere, characterized by exceptionally high connectivity with nearly all brain regions. Despite extensive animal studies on its anatomy and function and growing evidence of claustral deficits in neuropsychiatric disorders, its specific roles in normal and abnormal human brain function remain largely unknown. This is primarily due to its thin and complex morphology, which limits accurate anatomical delineation and neural activity isolation in conventional *in vivo* neuroimaging. To facilitate future neuroimaging studies, we developed a comprehensive and reliable mepanual segmentation protocol based on a cellular-resolution brain atlas and high-resolution ($0.7^3$ mm) MRI data. The protocols involve detailed guidelines to delineate the entire claustrum, including the inferior parts that have not been clearly described in earlier MRI studies. Additionally, we propose a geometric method to parcellate the claustrum into three subregions (the dorsal, ventral, and temporal claustrum) along the superior-to-inferior axis. The mean bilateral claustrum volume in 10 young adults was 3307.5 mm³, approximately 0.21% of total intracranial volume. Our segmentation protocol demonstrated high inter- and intra-rater reliability (ICC > 0.89, DSC > 0.85), confirming its replicability. This comprehensive and reliable manual segmentation protocol offers a robust foundation for anatomically precise neuroimaging investigations of the human claustrum.

**Introduction**

The claustrum is a thin deep-brain structure located in the basolateral telencephalon of the mammalian brain, present in all mammals with a cerebral cortex. Located at the center of each hemisphere, it is the brain's most highly connected hub, with reciprocal connectivity to nearly all brain regions and substantial input from neuromodulator circuits (Smythies et al., 2012; Torgerson et al., 2015). The claustrum has dense anatomical connections with cortical areas, including temporal, motor, somatosensory, visual, auditory, limbic, associative, sensorimotor, and prefrontal cortices (Brown et al., 2017; Goll et al., 2015; Milardi et al., 2015; Rodríguez-Vidal et al., 2024). Its extensive connectivity, particularly with the frontal cortex, enables feed-forward inhibition to cortical areas (Edelstein & Denaro, 2004; Jackson et al., 2020). The claustrum also has strong reciprocal connections with subcortical structures, including the basal ganglia and limbic system (Dillingham et al., 2017; Milardi et al., 2015; Rodríguez-Vidal et al., 2024). Animal studies implicate the claustrum in key brain functions such as multisensory integration (Mathur, 2014), conscious perception (Crick & Koch, 2005), and attentional control (Goll et al., 2015; White et al., 2018a; White & Mathur, 2018). It plays a crucial role in regulating slow cortical oscillations (0.5–4 Hz) that synchronize global cortical networks during mental state transitions (Narikiyo et al., 2020). Additionally, the claustrum has been linked to wakefulness (Atlan et al., 2024a; Lamsam et al., 2024), decision-making, cognitive control (Madden et al., 2022), network switching (Reser et al., 2014, 2017), and sleep regulation (Atlan et al., 2024b), primarily through its modulation of cortical activity and cortico-cortical communication.

Consistent with its hypothesized role in cognitive functions and brain network dynamics, growing evidence links the claustrum to various neurological disorders, including neurodevelopmental conditions like autism (Davis, 2008; Wegiel et al., 2015) and neurodegenerative diseases such as Alzheimer's (Bruen et al., 2008; Venneri & Shanks, 2014), Parkinson's disease (Arrigo et al., 2019; Kalaitzakis et al., 2009; Sitte et al., 2017), and Lewy body dementia (Kalaitzakis et al., 2009; Yamamoto et al., 2007). The claustrum's strong reciprocal connections with cortical, subcortical, and limbic regions suggest a key role in epilepsy, as lesions can eliminate or alter convulsive seizures in animals (Kudo & Wada, 1995; Majak et al., 2002; Mohapel et al., 2000; Wada & Kudo, 1997). Additionally, electrical stimulation studies indicate its involvement in altered consciousness and memory loss during seizures (Koubeissi et al., 2014). The claustrum has also been implicated in severe psychopathologies, including depression (Wang et al., 2023) and schizophrenia (Cascella et al., 2011a; Schinz et al., 2025). Evidence from cases of bilateral claustral lesions (Ishii et al., 2011; Sperner et al., 1996), post-mortem studies (Bernstein et al., 2016), a volumetric MRI study (Cascella et al., 2011b), and fMRI meta-analysis (van Lutterveld et al., 2013; Zmigrod et al., 2016) link claustral deficits to psychotic symptoms.

Although the claustrum is believed to play a crucial role in brain function and neuropsychiatric disorders (Meletti et al.,

2015; Silva et al., 2017), its study has been hindered by technical and anatomical challenges. Isolated bilateral claustrum lesions are extremely rare, limiting neuropsychological research. The structure's distinctive anatomy—an irregular, thin (1-5 mm) sheet-like formation deep within the brain—has made it particularly challenging to identify and delineate *in vivo* using conventional neuroimaging tools. As a result, the claustrum is absent from current brain neuroanatomic parcellation schemes, and no reliable method exists for comprehensive anatomical delineation. Additionally, functional neuroimaging methods (PET, fMRI, MEG, EEG) lack the spatial resolution to isolate claustral activity, preventing precise investigation of its role in cognition, behavior, and neuropsychiatric disorders.

The claustrum is described in several high-resolution histological atlases (Casamitjana et al., 2024; Ding et al., 2017; Ewert et al., 2018; Mai et al., 2015), and investigating the claustrum using neuroimaging has become possible with advances in high-resolution MRI. However, widely used probabilistic brain atlases (e.g., Harvard-Oxford cortical and subcortical brain atlas) and a brain atlas based on connectional architecture (Brainnetome atlas; http://atlas.brainnetome.org/index.html) do not include the claustrum. Therefore, the claustrum has been largely ignored in human neuroanatomical studies, and its potential roles in cognitive functions and neuropsychopathology have been misassigned to nearby structures (e.g., insula, putamen, amygdala, etc.) in functional neuroimaging studies.

While recent machine-learning based approaches have developed automatic brain regional segmentation tools including those delineating the claustrum (Albishri et al., 2022; Berman et al., 2020; Brun et al., 2022; Casamitjana et al., 2024; Mauri et al., 2024), their segmentation accuracy is limited and hard to be truly assessed without assessment of the reliability of the manually segmented claustrum labels that were used as the ground truth of the machine learning modeling. Prior manual training protocols (Coates & Zaretskaya, 2024; Davis, 2008) exist, but they lack sufficient detail on the claustrum's unique structure, graphical guidance, and clearly defined subregional boundaries. Furthermore, the reliability of these protocols was not assessed, limiting their usefulness.

Due to its thin structure and irregular sheet-like shapes, the claustrum is challenging to delineate using MRI. Its morphology varies due to surrounding white matter fibers' high plasticity (Bengtsson et al., 2005; Blumenfeld-Katzir et al., 2011). Therefore, there are substantial individual differences in morphology. In most mammals, it is broadly divided into dorsal and ventral parts, termed the "insular claustrum" and "endopiriform nucleus" in non-primate mammals (Kowiański et al., 1999; Reser et al., 2014). These regions differ in cell density, calcium-binding proteins, and connection profiles: the dorsal claustrum connects primarily with the cortex, while the ventral claustrum has strong connections with subcortical areas. Additionally, a distinct claustral subdivision termed the periamygdaloid claustrum has been described in human studies (Mai et al., 2015; Tyszka & Pauli, 2016), which roughly corresponds to the temporal subregion of the claustrum that is described in the most recent human brain atlas (Ding et al., 2017). Despite these prior efforts, a clear and standardized parcellation of the human claustrum remains lacking.

In our preliminary work (Kang, et al., 2020), we developed a comprehensive protocol for manually tracing the human claustrum. We used high-resolution (0.7 mm³ isotropic voxel size) T1-weighted MRIs from the Washington University-Minnesota Consortium Human Connectome Project (WU-Minn HCP; Van Essen et al., 2013a). The protocol was based on a cellular-resolution human brain atlas (Ding et al., 2017) that integrates neuroimaging (T1- and diffusion-weighted MRI), high-resolution histology, and large-format cellular-resolution (1 μm/pixel) Nissl and immunohistochemistry anatomical plates of a complete adult brain. It includes detailed descriptions for delineating three subregions of the claustrum: the dorsal, ventral, and temporal claustrum, which were identified based on the cytoarchitecture, chemoarchitecture, and connectivity of the brain atlas. The ventral and temporal claustrum in our protocol approximately correspond to the superior and inferior parts of the ventral claustrum identified by Fernández-Miranda et al. (2008).

In the present study, we assessed the reliability of the manual segmentation protocol by examining intra- and inter-rater reliability. Two operators manually segmented the whole claustrum in T1-weighted MRI datasets from 10 HCP human subjects, and a third operator and a custom computer program parcellated its subregions. Additionally, we measured claustrum volumes and created 3D images of the claustrum and its subregions, based on a rigorous and comprehensive protocol that includes the temporal claustrum. This work provides a clear depiction of the whole and subregional anatomy of the human claustrum, which has not been clearly demonstrated in prior studies.

**Methods**

*Subjects*

To develop the manual segmentation protocol for the claustrum, we randomly selected brain MRI scans of 10 human subjects (5 males, 5 females, age range 22–35 years), with age ranges matched between the two sexes. from the WU-Minn HCP database. The HCP subjects were drawn from a population of young, healthy adults (ages 22–35) without any prior history of significant neurological and psychiatric illnesses, which were determined based on semi-structured interviews and comprehensive neurological and psychological assessments. Further details on the HCP subjects are described elsewhere (David C Van Essen et al., 2012; David C Van Essen et al., 2013).

*MRI scan acquisition and processing*

All subjects were scanned on a customized Siemens 3T Connectome Skyra scanner system (a Siemens SC 72 gradient coil and standard 32-channel Siemens head coil). Structural images were acquired using the 3D MPRAGE T1-weighted sequence with 0.7 mm isotropic resolution (FOV=224 mm, matrix=320, 256 sagittal slices in a single slab, TR=2400 ms, TE=2.14 ms, TI=1000 ms, flip angle=8°). The acquisition methods have been extensively described in van Essen et al. (2012b). The T1-weighted MRI data was preprocessed according to the HCP protocol, as described elsewhere (Glasser et al., 2013).

*Segmentation protocol*

We used 3D Slicer software (http://www.slicer.org; version 4.10.1) and the web-based cellular-level Allen Brain Atlas of an adult human (http://atlas.brain-map.org/; Ding et al., 2017), which offers high user interactivity for zooming, highlighting specific annotation regions, and more. Additionally, we used the BigBrain images scanned at 20 μm isotropic resolution (Amunts et al., 2013) for cross-referencing with the Allen Brain Atlas. The brain atlas provides structural annotations for 862 brain regions across 106 coronal plates, sectioned at an oblique angle to the AC-PC plane of WU-Minn HCP T1-weighted MRI data. The anterior temporal claustrum, among the most challenging regions to trace, is visible in Slab 4 slices, sectioned at an ~10-degree oblique angle (Figure 16 of Ding et al., 2017). During the development of the MRI segmentation protocol, we generated rotated MRI volumes by re-gridding with linear interpolation using the Slicer Crop Volume module to adapt anatomical landmarks from the atlas to HCP MRIs. However, for the present study, we conducted manual segmentation using the original AC-PC aligned MRIs, as rotation did not significantly improve segmentation accuracy. To ensure equivalent contrast range across the claustrum segmentations, T1 images were adjusted by setting the display window width and level (brightness and contrast) to 500 and 700, respectively, which provided the best visibility of the claustrum nuclei.

A manual tracing approach for subcortical structures (e.g., amygdala, hippocampus, and caudate) primarily relies on the coronal planes (Entis et al., 2012; Hashempour et al., 2019; Moore et al., 2014; Morey et al., 2008). However, the claustrum has a thin and complex morphology that is shaped by adjacent white matter and gray matter structures. Therefore, it requires a more sophisticated tracing strategy that incorporates different planes for distinct subregions. For example, the dorsal claustrum is most easily traced in the axial planes, whereas the ventral claustrum is best traced in the coronal planes. Relying on a single-plane perspective to trace the entire claustrum can easily result in errors, such as the inclusion of adjacent cortical and subcortical voxels from transitional areas that appear connected to the claustrum. Therefore, it is strongly recommended to trace the claustrum in the specified order of planes, as detailed below. It is also critical to correct any tracing errors through a final visual inspection across all three views.

*Tracing the dorsal claustrum*

Tracing begins in the axial plane, starting near the inferior aspect of the central anterior commissure (AC), where the thin morphology of the dorsal claustrum, located between the putamen and the insular cortex, is clearly visible (see Figure 1). The dorsal claustrum is elongated along the anterior-posterior axis and medially separated from the putamen by the external capsule and laterally from the insular cortex by the extreme capsule. The rostral, caudal, and dorsal boundaries of the dorsal claustrum approximately align with those of the putamen. The dorsal claustrum narrows superiorly; therefore, tracing should continue dorsally until the small dorsal claustrum nuclei near the dorsal boundary are no longer distinguishable. Tracing the dorsal extremities of the dorsal claustrum should be completed in the subsequent tracing of the ventral claustrum in the coronal planes, which provide a clearer view of the dorsal boundaries.

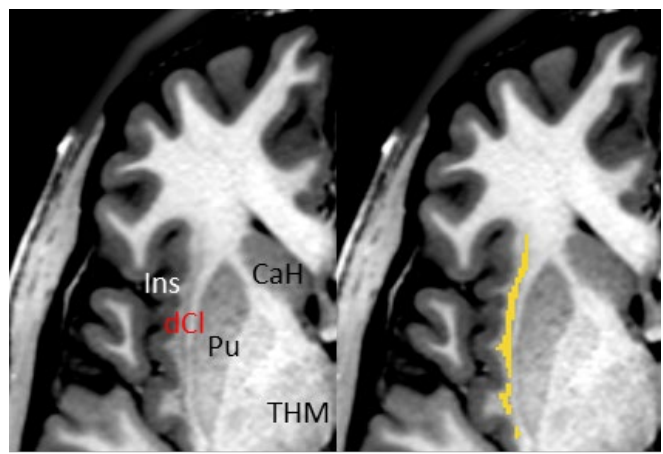

**Figure 1.** The dorsal claustrum (dCl) in the axial planes. dCl lies between the putamen (Pu) and the insular cortex (Ins). The external and extreme capsules separate dCl from Pu and Ins, respectively.

*Tracing the ventral claustrum*

The ventral claustrum is clearly defined in the coronal view, where it is distinctly separated from surrounding structures by the external and extreme capsules (see Figure 2). Begin tracing from the coronal plane that intersects the midpoint of the claustrum in the axial view. Proceed anteriorly until the claustrum nuclei are no longer distinguishable. Then, return to the starting point to complete tracing the posterior ventral claustrum. While Figure 2 depicts the entire claustrum in the coronal view, at this stage, trace only the portion of the ventral claustrum located superior to the most inferior medial point of the putamen (see Figure 4 below). In the coronal view, the anterior ventral claustrum extends inferior to the putamen, following its external curvature. The anterior ventral claustrum is distinctly separated dorsomedially from the putamen by the external capsule and ventrolaterally from the base of the frontal lobe, including the orbitofrontal cortex and frontal agranular insular cortex, by the extreme capsule. However, the inferior border of the anterior ventral claustrum partially adjoins orbitofrontal cortex (OFC) and FI (frontal agranular insular cortex) (Figure 2, rows ii and iii). Therefore, careful tracing of the anterior ventral claustrum is required to avoid the erroneous inclusion of adjacent frontal regions near its inferior border. They are distinguishable in the coronal planes. The rostral and caudal borders of the ventral claustrum are more clearly identified in the axial view; therefore, they should be finalized during the final visual inspection in the axial plane.

*Tracing the temporal claustrum*

The inferior portion of the ventral claustrum extends into the temporal lobe, forming the temporal claustrum. Along the anterior-to-posterior axis, it extends from the rostral tip of the endopiriform nucleus (EN) to the rostral tip of the hippocampus (see Figure 2 rows iv and vii which show the rostral and the caudal ends of the temporal claustrum). The anterior portion of the temporal claustrum is the most challenging region of the claustrum to trace because it is surrounded by the gray matter of frontal and temporal lobes and adjoins the amygdaloidal complex, particularly EN. Due to its close relationship with the amygdaloid complex, the anterior temporal claustrum is often referred to as the periamygdalar claustrum (Zelano & Sobel, 2005a). The ventrolateral border of the anterior temporal claustrum adjoins the temporal angular insular cortex, the entorhinal cortex (EC; in the rostral part; see Figure 2, rows iv and v) and white matter bundles, including the uncinate fasciculus and the inferior fronto-occipital fasciculus (in the caudal part). The dorsomedial border of the anterior temporal claustrum adjoins EN and the piriform cortex. Due to multiple transitional areas between the anterior temporal claustrum and the adjacent regions, clearly delineating its borders is challenging. Despite its complex anatomy, the anterior temporal claustrum can be identified on high-resolution MRIs (0.7 mm³ isotropic voxel size) due to its nuclei appearing darker than the surrounding gray matter, although delineating its borders remains challenging. As shown in the Nissl-stained brain slices in Figure 2, the temporal claustrum contains dense gray matter nuclei that stand out against the background of adjacent cortical and subcortical structures in the medial temporal lobe.

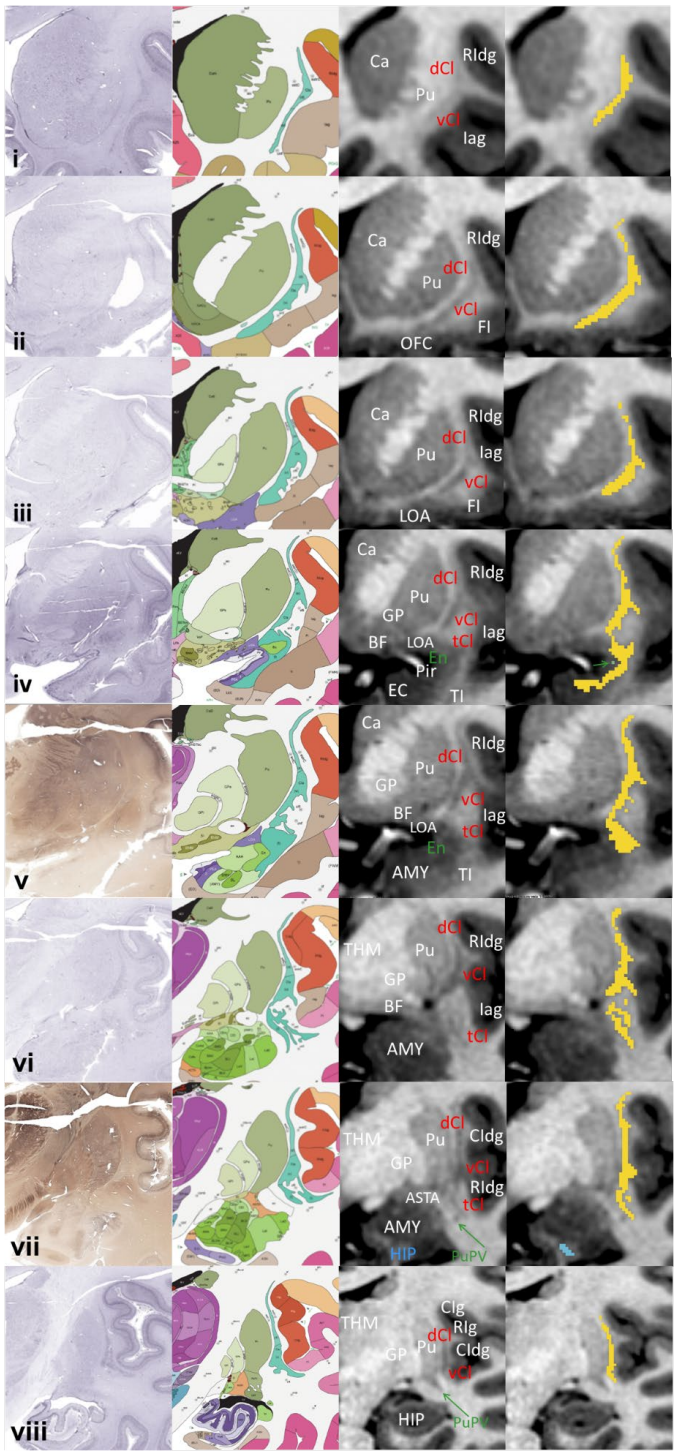

**Figure 2.** Comparison of histological photographs from the cellular resolution brain atlas (Ding et al., 2017) with MR images and tracings on one young subject at different levels, from most rostral (row i) to most caudal (row viii) boundaries of the claustrum (Cl). From left to right, columns show histological images, the atlas parcellation labels, MRI slices, and whole claustrum tracings in yellow. Postmortem histological sections are shown to provide a standard set of reference images to illustrate the landmarks in our protocol. The labels of the three subregions of Cl (dorsal, ventral, and temporal Cl; dCl, vCl, and tCl) and the landmarks described in the in the labeling protocol are depicted in the third column. Abbreviations: ASTA, amygdalostriatal transition area; BF, basal forebrain; Ca, caudate nucleus; CaH, caudate nucleus; Cl, the claustrum; dCl, dorsal claustrum; vCl, ventral claustrum; tCl, temporal claustrum; CIdg, caudal dysgranular insular cortex; EC, entorhinal cortex; En, endopiriform nucleus; FI, frontal agranular insular cortex; GP, globus pallidus; Iag, agranular insular cortex; Ins, insular cortex; LOA, lateral olfactory area; PI, parainsular cortex; Pir, pi-riform cortex; Pu, putamen; PuPV, posteroventral putamen; Ridg, rostral dysgranular insular cortex; OFC, orbitofrontal cortex; TI, temporal angular insular cortex.

In tracing the anterior temporal claustrum, both the coronal and axial planes should be used simultaneously, with primary tracing in the coronal planes followed by adjustments and confirmations in the corresponding axial planes. The coronal view is useful for identifying overall morphology, while the axial view aids in more accurately defining the borders (see Figure 3 rows i and ii). This procedure requires a side-by-side display of both planes and the use of a crosshair navigation tool to show intersections between them. Extra care must be taken to include only the anterior temporal claustrum, which appears darker than the surrounding gray matter. As the first step, EN should be identified as a key landmark for delineating the medial border of the anterior temporal claustrum in the coronal view. The rostral tip of EN is located near the junction of the frontal and temporal lobes (marked by a green arrow in Figure 2 row iv). Once the EN landmark is localized in the coronal plane, begin to trace the anterior temporal claustrum that adjoins the dorsolateral side of the landmark point and extends ventromedially. Continue tracing toward the posterior temporal claustrum, which is embedded within the white matter bundles of the temporal lobe.

The posterior temporal claustrum is relatively easier to trace in the coronal planes compared to the anterior temporal claustrum. However, the white matter fibers of the uncinate fasciculus and the inferior fronto-occipital fasciculus traverse the

## Parcellation of the claustrum into the three subregions

As described above, the claustrum is divided into three subregions along the superior-to-inferior axis. A geometric method can be applied to parcellate them accurately. In our earlier work, we developed a method utilizing three anatomical landmark points, which parcellated the claustral subregions effectively (Kang, et al., 2020). We further developed the method, simplifying the procedures to obtain more reliable parcellation of the claustrum. A unique aspect of our approach is the use of geometric landmarks derived from the neighboring putamen rather than the claustrum itself. Although this may appear unconventional, it is well-justified anatomically. The claustrum is a thin, elongated nucleus situated between the putamen and insula, and it follows the curvature of the outer surface of the putamen. Notably, consistent and easily identifiable anatomical landmarks useful for parcellation are more reliably observed in the putamen than in the claustrum itself.

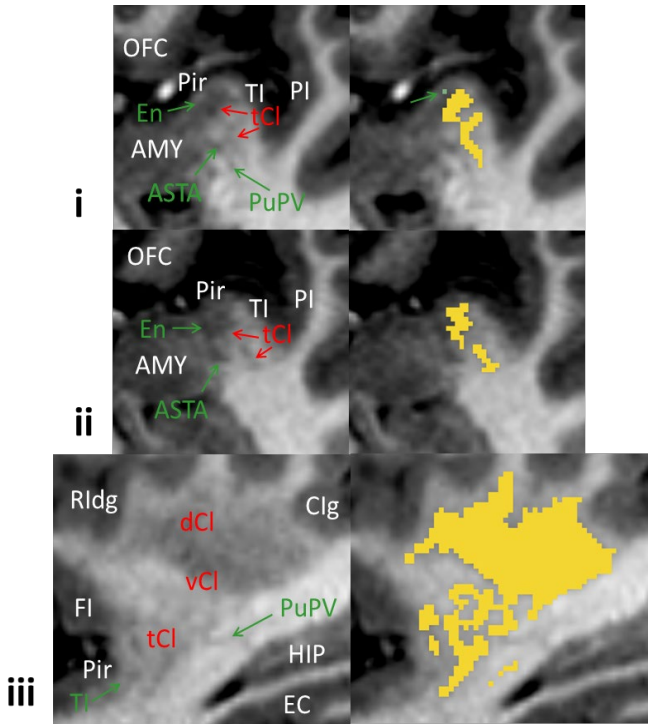

**Figure 3.** The axial planes (row i and ii) and the sagittal plane (row iii) that should be simultaneously used with the corresponding coronal planes for clear delineation of the temporal claustrum (tCl). The endopiriform nucleus (En) is the key landmark point useful for determining the medial border of tCl, which is highlighted with a green arrow and a green point (top left). The amygdala (AMY) and the piriform cortex (Pir) also adjoins to the medial border of tCl. Other nearby gray matter structures that can be easily confused with tCl, such as the amygdalostriatal transition area (ASTA), posteroventral putamen (PuPV), and temporal angular insular cortex (TI), are highlighted with green arrows.

posterior temporal claustrum, fragmenting its inferolateral portions into many small pieces, referred to as "puddles" (Johnson et al., 2014; Mathur, 2014). Therefore, tracing in the coronal view should be supplemented with the sagittal view, which clearly shows the main body of the lateral posterior temporal claustrum (see Figure 3, row iii). Despite the puddles, the most temporal part of the claustrum remains well connected as a whole, as shown in the 3D plot (Figure 5). Notably, sagittal planes should not be used to refine the medial border of the temporal claustrum, as distinguishing the transition from its medial border to the lateral amygdala is difficult in this view.

A final visual check using multiple planes is required for necessary corrections. Figure 3 shows the axial and sagittal planes, where the brain regions adjoining the temporal claustrum—EN, the amygdalostriatal transition area (ASTA), and the temporal angular insular cortex (TI)—are depicted in green. It also highlights adjacent areas that are difficult to differentiate from the temporal claustrum, including the posteroventral putamen (PuPV) in the coronal planes. It is also recommended to consult the online brain atlas (the "Adult Human" atlas at http://atlas.brain-map.org/) for additional guidance on anatomical details of the claustrum not fully described in the current protocol.

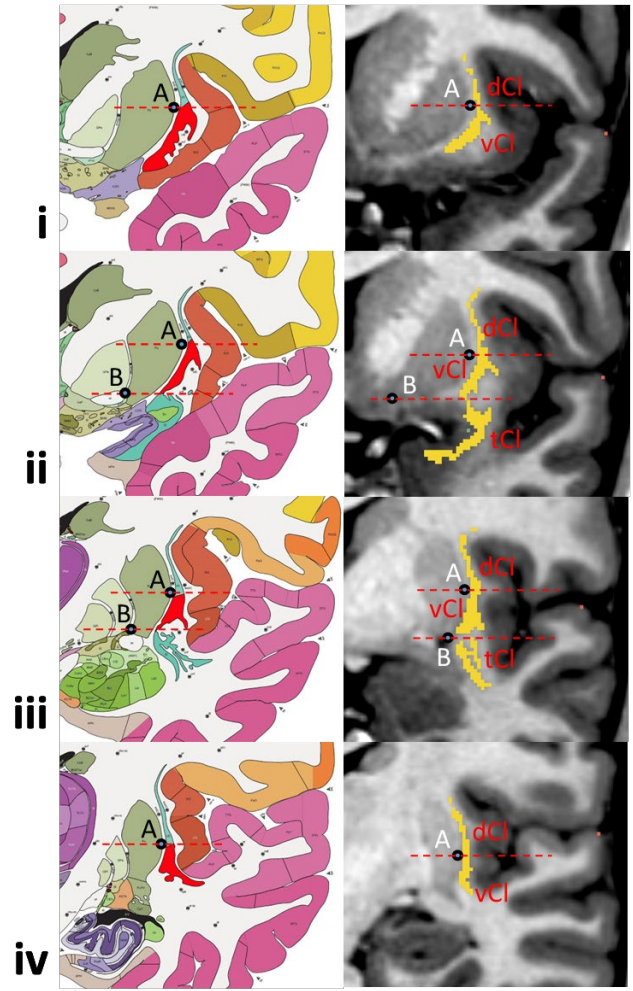

**Figure 4.** Detail of geometrically determined protocol for segmenting the three subregions of the claustrum (Cl). See text for the details of the protocol, including the explanation of points A and B. The dorsal and ventral Cl (dCl and vCl) appear in the entire coronal planes of Cl, while the temporal Cl (tCl) is observed only in the middle coronal planes (row ii and iii). vCl is highlighted with red color in the brain atlas parcellation label maps (left column).

Step-1 is to manually identify two key landmark points in the coronal planes manually (Figure 4): Point A, the most lateral point along the outer curvature of the putamen, located at approximately half its vertical height; Point B, the most inferior-medial point of the putamen. Points A should be identified across all coronal planes where the ventral claustrum is present. Point B should be identified only in the coronal planes spanning the anterior-to-posterior range of the temporal claustrum, which ranges from the coronal plane showing the rostral tip of EN (around 1-2 mm anterior to the rostral tip of the amygdala) to the coronal plane showing the posterior end of the amygdala. Step-2 is to draw two straight lines, including Line 1 extending horizontally from Points A and Line 2 extending horizontally from Point B. All parts of the claustrum above Line 1 are classified as the dorsal claustrum, all parts below Line 2 as the temporal claustrum, and the remaining region as the ventral claustrum. Step-2 is implemented programmatically using a custom program written in Matlab. Step-3 is to visually inspect and manually correct any obvious parcellation errors.

In addition to the semi-automatic method, we also developed a fully automated method where Step-1 is implemented using FreeSurfer (Khan et al., 2008) brain parcellation labels (*aparc+aseg.mgz*). A custom Matlab program automatically identifies the two landmark points and the anterior-posterior limits following the same anatomical criteria as the manual method and using FreeSurfer labels of the putamen and amygdala. As previous studies reported, FreeSurfer can erroneously segment the putamen with an "expansion of the lateral aspect to include parts of the external capsule and claustrum" (Dewey et al., 2010). To avoid errors in identifying point A caused by this issue, the program determines the height of Point A based on the vertical extent of the putamen, which aligns well with manually defined values. Then the program implements Step-2, parcellating the claustrum into its three subregions using the landmark points. In the fully automated method, Step-3 (parcellation error correction) was not performed.

*Volume Measurement*

We measured the volumes of the whole and subregional claustrum in each hemisphere by counting the number of voxels in each region and multiplying by the voxel volume (0.343 mm³, derived from 0.7 mm × 0.7 mm × 0.7 mm). In addition to the absolute volumes, we also calculated volumes adjusted for intracranial volume (ICV), which were obtained by dividing each volume by the total ICV estimated by FreeSurfer. ICV-adjusted, relative volumes account for individual differences in head size, enabling more accurate comparisons across subjects by normalizing regional volumes relative to overall brain size.

*Reliability Assessment*

Two operators (JB and KM) independently traced the entirety of the whole claustrum using the protocol described above to assess the inter-rater reliability. Additionally, to assess intra-rater reliability, one of the operators (JB) retraced the claustrum in all subjects. Parcellation of the claustral subregions requires two landmark points on the putamen, which can be identified independently of the claustrum anatomy. Thus, a third operator (SK) and the FreeSurfer parcellation-based program identified the landmark points manually and automatically, respectively. The third operator identified the landmark points twice to assess intra-rater reliability. For each set of landmark points, the custom program parcellated the three claustral subregions. We assessed the reliability of the subregion parcellation protocol using the first whole claustrum tracing sets from the first operator, which allows controlling for the confounding effect of inter-rater variation in the whole claustrum tracings (Entis et al., 2012).

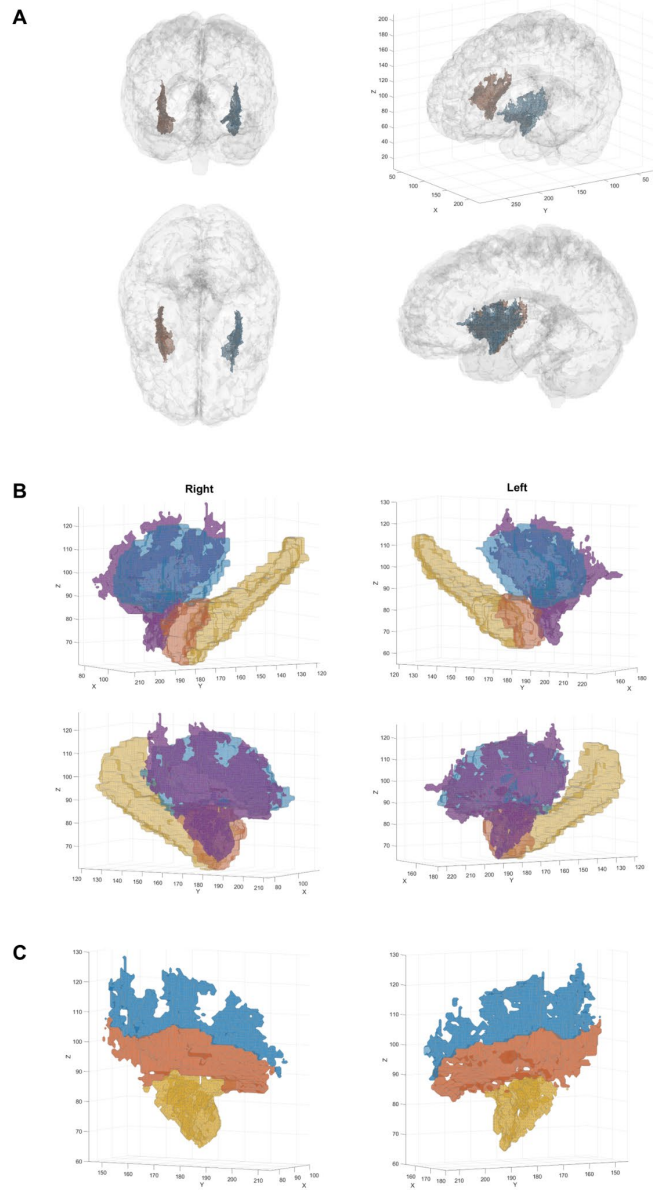

**Figure 5.** The 3D models of the manually segmented claustrum. **A.** The whole claustrum (blue: left; red: right) in 3D space. **B.** The whole claustrum (purple) in the left and right hemisphere depicted with the adjacent subcortical structures: the putamen (blue), amygdala (orange), and hippocampus (yellow). **C.** The 3D models of the dorsal (blue), ventral (red), and temporal (yellow) claustrum depicted in the right and left hemisphere spaces.

**Table 1.** Mean absolute and relative volumes of the whole claustrum and its subregions in both hemispheres. Absolute volumes are in mm³, and relative volumes are % of claustral to total intracranial volume. Standard deviations are in parentheses.

|  |  | Operator JB | | Operator KM | | Operator JB | |
|---|---|---|---|---|---|---|---|
|  |  | Left | Right | Left | Right | Left | Right |
| Whole | Absolute | 1693.0 | 1729.0 | 1637.1 | 1673.7 | 1701.7 | 1757.3 |
|  | (SD) | (239.3) | (273.3) | (266.9) | (308.2) | (260.3) | (301.1) |
|  | Relative | 0.104 | 0.105 | 0.099 | 0.102 | 0.104 | 0.107 |
|  | (SD) | (0.016) | (0.015) | (0.013) | (0.015) | (0.015) | (0.016) |
|  |  | Operator SK | | Program using FreeSurfer | | Operator SK (retrace) | |
|  |  | Left | Right | Left | Right | Left | Right |
| Dorsal | Absolute | 483.5 | 465.7 | 463.7 | 450.7 | 480.6 | 460 |
|  | (SD) | (141.2) | (136.1) | (130.3) | (119.8) | (114.7) | (106.0) |
|  | Relative | 0.029 | 0.028 | 0.028 | 0.027 | 0.029 | 0.028 |
|  | (SD) | (0.008) | (0.007) | (0.008) | (0.007) | (0.007) | (0.006) |
| Ventral | Absolute | 831.7 | 816.1 | 854.7 | 813.7 | 835.6 | 822.1 |
|  | (SD) | (133.3) | (139.0) | (133.2) | (132.1) | (127.2) | (135.9) |
|  | Relative | 0.051 | 0.050 | 0.052 | 0.050 | 0.051 | 0.050 |
|  | (SD) | (0.010) | (0.010) | (0.009) | (0.009) | (0.009) | (0.009) |
| Temporal | Absolute | 377.8 | 447.2 | 374.6 | 464.6 | 376.8 | 446.9 |
|  | (SD) | (118.5) | (152.5) | (95.0) | (148.0) | (103.6) | (145.5) |
|  | Relative | 0.023 | 0.027 | 0.023 | 0.028 | 0.023 | 0.027 |
|  | (SD) | (0.008) | (0.008) | (0.007) | (0.008) | (0.007) | (0.008) |

For the reliability assessments, we calculated two volumetric agreement measures: the intraclass correlation coefficient (ICC) (Shrout & Fleiss, 1979) and Dice similarity coefficient (DSC) (Dice, 1945). The ICC, specifically a two-way mixed-effects model with absolute agreement, quantifies the consistency and agreement of volumetric measurements between operators, accounting for systematic differences. In contrast, the DSC evaluates the spatial overlap between two segmentations as the ratio of twice the number of intersecting voxels to the sum of the voxel counts in both segmentations. We used MATLAB version 2023a (MathWorks Inc.) for all the quantitative analysis.

## Results

*Morphology of the claustrum*

Figure 5 shows 3D models of the manually segmented bilateral claustrum and its subregions in an individual (HCP ID: 164636), traced by a single operator. As illustrated in the top 3D plot, the claustrum is thin yet represents a substantially large subcortical nucleus, with an anterior-posterior extent comparable to that of the putamen and a height equivalent to the combined height of the putamen and amygdala. The three subregions are parcellated along the superior-to-inferior axis, and their 3D models illustrate the unique shape of the claustrum (Figure 5C). The subdivisions of the claustrum closely surround adjacent subcortical structures: the dorsal claustrum covers the superior-lateral surfaces of the putamen, the ventral claustrum covers the lateral and ventrolateral surfaces of the putamen, and the temporal claustrum extends inferiorly over the amygdala.

*Volumes of the claustrum*

The volumes of the whole and subregional claustrum in the left and right hemisphere are summarized in Table 1. The mean volumes of the left and right claustrum, averaged across the 10 individuals and the two operators, were 1677.3±236.5 mm³ and 1720.0±276.2 mm³, which correspond to 0.102±0.014 % and 0.105±0.014 % of the total intracranial volume, respectively. The right claustrum was slightly larger than the left claustrum, although the hemispheric difference was not statistically significant in a paired *t*-test with this small sample (absolute volume: $t(9) = 1.32$, $p = .220$, Cohen's $d = -0.42$; relative volume: $t(9) = 1.16$, $p = 0.276$, Cohen's $d = -0.37$). An independent *t*-test indicated that the absolute claustral volumes of the five males were somewhat larger than those of the five females, particularly in the right hemisphere (1891.6±212.7 mm³ vs. 1584.4±229.7 mm³; $t(8) = 2.45$, $p = 0.040$, Cohen's $d = 1.55$) rather than left hemisphere (1785.1±213.8 mm³; vs. 1569.4±225.9 mm³; $t(8) = 1.55$, $p = 0.160$, Cohen's $d = 0.98$). However, the relative volumes, adjusted for the individual differences in the intracranial volume, were comparable between males and females in both hemispheres (right: 0.107±0.017 % vs. 0.106±0.012 %; $t(8) = 0.47$, $p = 0.654$, Cohen's $d = 0.29$; left: 0.100±0.018 % vs. 0.104±0.010 %; $t(8) = -0.34$, $p = 0.743$, Cohen's $d = -0.21$).

*Reliability of the Manual Segmentation of the Claustrum*

The ICC and DSC measures of manual segmentation reliability are presented in Table 2. For the whole claustrum segmentation, both reliability metrics demonstrated excellent performance. The inter-rater reliability of the claustrum segmentation was excellent, with an ICC of 0.898 bilaterally, 0.863 for the left hemisphere, and 0.919 for the right hemisphere. The intra-rater reliability was even slightly higher, with ICC values consistently above 0.90 (bilateral: 0.906; left: 0.904; right: 0.908). Similarly, high DSC values were achieved for both inter-rater (bilateral: 0.838±0.025; left: 0.837±0.027; right: 0.839±0.028) and intra-rater reliability (bilateral: 0.869±0.032; left: 0.869±0.033; right: 0.870±0.035), confirming the protocol's for claustral subregion parcellation exceeded those of the whole structure. The semi-automatic and automatic parcellation methods showed strong agreement, with bilateral subregions achieving exceptionally high average ICC and DSC values of 0.959 and 0.932, respectively. The intra-rater reliability of semi-automatic parcellation was similarly high, with average ICC and DSC values of 0.962 and 0.938 for bilateral subregions. The consistently high reliability across all subregions validates the robustness of our parcellation protocol.

**Table 2.** Reliability of the manual segmentation labels of the whole and subregional claustrum measured by intraclass correlation coefficient (ICC) and Dice similarity coefficient (DSC). The inter-rater reliability for the subregion parcellation was assessed by comparing the semi-automated and fully automated methods. The standard deviations of the DSC are reported in parentheses.

|  |  | Inter-rater Reliability | | | Intra-rater Reliability | | |
| --- | --- | --- | --- | --- | --- | --- | --- |
|  |  | Bilateral | Left | Right | Bilateral | Left | Right |
| Whole | ICC | 0.898 | 0.863 | 0.919 | 0.906 | 0.904 | 0.908 |
|  | DSC | 0.838 | 0.837 | 0.839 | 0.869 | 0.869 | 0.870 |
|  | (SD) | (0.025) | (0.027) | (0.028) | (0.032) | (0.033) | (0.035) |
| Dorsal | ICC | 0.921 | 0.925 | 0.921 | 0.922 | 0.925 | 0.924 |
|  | DSC | 0.925 | 0.924 | 0.925 | 0.931 | 0.929 | 0.933 |
|  | (SD) | (0.011) | (0.014) | (0.013) | (0.010) | (0.012) | (0.010) |
| Ventral | ICC | 0.973 | 0.968 | 0.964 | 0.975 | 0.971 | 0.976 |
|  | DSC | 0.932 | 0.933 | 0.932 | 0.940 | 0.941 | 0.938 |
|  | (SD) | (0.012) | (0.016) | (0.011) | (0.008) | (0.010) | (0.010) |
| Temporal | ICC | 0.982 | 0.949 | 0.992 | 0.988 | 0.985 | 0.991 |
|  | DSC | 0.938 | 0.945 | 0.948 | 0.944 | 0.957 | 0.954 |
|  | (SD) | (0.015) | (0.023) | (0.025) | (0.008) | (0.019) | (0.015) |

## Discussion

Using recent advancements in cellular-resolution human brain atlases and sub-millimeter high-resolution MRI, we developed protocols for manual segmentation of the human claustrum. The protocol provides comprehensive step-by-step guidelines for tracing the claustrum in axial, coronal, and sagittal planes, along with final visual checks and corrections using multi-plane views to ensure accurate segmentation of this thin, elongated, irregularly shaped structure. Additionally, unlike previous studies that recognized only two subregions (dorsal and ventral) of the claustrum, our protocol further parcellates the claustrum into three subregions along the superior-to-anterior axis (Figure 5) based on an atlas that differentiates subcortical nuclei according to the cytoarchitecture and chemoarchitecture of the neurons (Ding et al., 2017). The inter-rater and intra-rater reliability measures of the bilateral claustrum manual segmentation were very high, with ICC values of 0.898 and 0.906 and DSC values of 0.838 and 0.869, respectively. According to the evaluative criteria for ICC (0.50-0.75: moderate; 0.75-0.90: good; > 0.90: excellent; Koo & Li, 2016), our protocol has high reliability, allowing consistent manual segmentation. Although there are no universally agreed-upon evaluative criteria for DSC, which requires context-based interpretation and tends to be low for small, complex, or thin structures (Maier-Hein et al., 2024), the DSC 0.838-0.869 is also considered excellent agreement in MRI segmentation. Furthermore, the inter-rater reliability between the semi-automatic and fully automatic parcellations (ICC: 0.971; DSC: 0.932) and the intra-rater reliability of the semi-automatic parcellations (ICC: 0.967; DSC: 0.933) were even higher. To control for the confounding effect of inter-rater variability in whole-claustrum tracing on the subregional parcellation, we used the same whole-claustrum tracing set from a single operator for reliability assessment. While this factor may have contributed to the high reliability, the more critical advantage of our method lies in its geometric approach. Specifically, it identifies two landmark points using a clearly visible subcortical structure (i.e., the putamen) and applies an automated procedure to parcellate the subregions based on these landmarks. This straightforward approach facilitated strong reproducibility in the parcellation outcomes.

We found that the total volume of the bilateral claustrum was 3307.5 mm³ (left: 1693.0±239.3 mm³; right: 1729.0±272.3 mm³), corresponding to about 0.21% of the total intracranial volume. It is notable that the claustral volumes are comparable to the amygdala volume manually segmented from high-resolution MRI (left: 1727.6±177.9 mm³; right: 1750.3±218.9 mm³; Entis et al., 2012), indicating that the claustrum is a sizeable subcortical structure. To our knowledge, eight studies—including ours—have reported human claustrum volumes (Table 3). Six of these used post-mortem brains: four employed histological sections and two used ultra-high-resolution *ex vivo* MRI. The remaining two studies used *in vivo* MRI. There are notable differences across studies, part of which might be due to different imaging resolutions and modalities. Since claustrum volume varies with total intracranial volume and age (Bennett & Baird, 2006), the absolute volumes reported across studies may not be directly comparable, as most did not provide relative volume measures. More importantly, the lack of detailed manual segmentation protocols in several studies likely contributed to variability in reported volumes—particularly the lower values found in older studies. Interestingly, reported claustrum volumes appear to increase over time and converge toward values similar to those found in the present study. This trend may reflect improved identification of the anatomical boundaries of the human claustrum, facilitated by advances in high-resolution imaging and 3D visualization technologies.

Table 3. Claustrum volumes reported in the previous studies and the present study. Volumes are reported as the average of males and females. Reporting formats for sex, age, laterality, and volume (e.g., mean ± SD, individual values, or ranges) differ across studies due to variations in sample size and publication reporting practices.

| Study | Dataset | Subjects | Volume (mm³) |
|---|---|---|---|
| Kowiański et al. (1999) | Histological brain sections | 5 humans (sex and age unknown) | 580.0 ± 58.0 (laterality unknown) |
| Kapakin (2011) | Histological brain sections | 1 male (age unknown) | L: 705.8 R: 828.8 |
| Milardi et al. (2015) | *In vivo* T1 MRI (1.0³ mm) | 5 males & 5 females (mean 32.1 years) | L: 804.0 (range: 752.0–912.0) R: 813.6 (range: 744.0–864.0) |
| Bernstein et al. (2016b) | Histological brain sections | 9 males & 6 females (47.3 ± 7.6 years) | L: 958.1 ± 235.9 R: 941.0 ± 228.8 |
| Carlarco et al. (2023) | BigBrain (0.1 mm³ 3D histology) | 1 male (65 years) | L: 1243.3 (no report on the right claustrum) |
| Coates & Zeretskaya (2024) | *Ex vivo* T1 MRI (0.1 mm³) | 1 female (58 years) | L: 1736.2 R: 2074.0 |
| Mauri et al. (2024) | *Ex vivo* T1 MRI (0.1 mm³) | 9 males & 7 females (63.9 ± 12.9 years) | 1253.05 ± 283.79 (mean of the left and right claustrum) |
| Present study | *In vivo* T1 MRI (0.7 mm³) | 5 males & 5 females (22 – 35 years) | L: 1693.0 ± 239.3 R: 1729.0 ± 272.3 |

The volumes of the three subregions were disproportionate: the volumes of the dorsal, ventral, and temporal claustrum corresponding to 0.27:0.51:0.22 in the left claustrum while 0.26:0.48:0.26 in the right claustrum. Roughly speaking, the ventral claustrum consists of about 50% of the whole claustrum, while the dorsal and temporal claustrum each consist of about 25% of the whole claustrum. It is the first report of the three subregions of the claustrum and their volumes in the human brain. The dorsal and ventral claustrum are the primary claustrum structures described in MRI studies, being positioned between the putamen and insula. The temporal claustrum is referred to as the periamygdalar claustrum due to its proximity to the amygdaloid complex (Zelano & Sobel, 2005b). As Figure 5 illustrates, the temporal claustrum extends along the ventral side of the amygdala. The temporal claustrum has been largely excluded in MRI studies, probably due to limited spatial resolutions of the conventional MRI scans hindering clear identification of the structure (Arrigo et al., 2017; Milardi et al., 2013; Torgerson et al., 2015).

Given the proximity to the different subcortical and cortical structures, the three subregions of the claustrum may be involved in similar but distinct functions. For example, the temporal claustrum, which is in close anatomical proximity to the amygdala, might be related to the amygdala's function as a hub of the emotional salience network (Ince et al., 2023; Kong & Zweifel, 2021). While the limited spatial resolution of conventional fMRI precludes clear differentiation of claustrum activity from that of adjacent regions, consistent reports of periamygdala activation in response to emotionally salient stimuli (Kilts et al., 2003; Pannu Hayes et al., 2009) may partly reflect BOLD activity attributable to the temporal claustrum, suggesting its potential involvement in emotional salience processing. In contrast, given the position and the animal study findings of the claustrum's close collaboration with anterior cingulate cortex for top-down control of behavior (White et al., 2018b), the dorsal claustrum may have significant role in attentional control. As the central main body of the claustrum positioned between the putamen and insula, the ventral claustrum appears to be a part of the cortico-basal ganglia circuitry (Borra et al., 2024), playing a role in salience processing and motor control. As a growing body of functional neuroimaging studies have revealed additional roles of the claustrum (Coates et al., 2024; Rodríguez-Vidal et al., 2024; Stewart et al., 2024), future studies should systematically investigate its functions using various experimental paradigms and large samples.

The present manual segmentation protocol serves multiple important purposes in neuroimaging research. It promotes a deep understanding of neuroanatomy, helps recognize anatomical variability, and functions as a valuable educational tool for training researchers and clinicians. Manual segmentation is also essential for handling atypical anatomy, pathological cases, high-stakes clinical decisions, and novel research questions.

While automated methods are becoming increasingly prevalent, manual protocols remain fundamental for ensuring reliability and advancing the field. For example, recent studies have developed deep learning–based tools for automatic claustrum segmentation (Albishri et al., 2022; Li et al., 2021). These tools reduce the need for labor-intensive efforts and enable large-scale *in vivo* MRI studies. However, their success depends heavily on the quality of MRI data and the availability of accurate manual segmentations, which serve as ground truth for training and validation. Our reliable and comprehensive manual segmentation protocol will help establish benchmarks for evaluating automated methods, identifying biases, and guiding continued improvements in claustrum segmentation algorithms.

*Limitations:* Using $0.7^3$ mm high-resolution, high-contrast T1-weighted MRI, we reliably delineated the entire claustrum by distinguishing its voxels from adjacent white and gray matter. However, our study has several limitations. First, this resolution is insufficient for accurately identifying the fragmented nuclei of the temporal claustrum, or "puddles" (Johnson et al., 2014; Mathur, 2014), which require ultra-high resolutions ($< 0.5^3$ mm voxel size; Coates & Zaretskaya, 2024). As a result, our segmentation may have included small interspersed white matter and missed some isolated nuclei. Both users of our protocol and automated segmentation algorithms based on it should clearly recognize the limitation in delineating the puddles in the temporal claustrum and the superior tip of the dorsal claustrum. However, visual inspection of the Allen Brain Atlas (Ding et al., 2017) and the BigBrain model (20 $\mu m$ isotropic; Amunts et al., 2013) suggests these excluded nuclei constitute only a minimal portion of the claustrum. Since *in vivo* MRI at such ultra-high resolutions remains challenging, precise investigation of these tiny "puddles" nuclei should rely on postmortem studies. Second, our segmentation protocols were developed using submillimeter MRIs, so applying them to lower-resolution MRIs ($>1.0$ mm³) may reduce accuracy, particularly in the anterior temporal and superior dorsal claustrum, where the periamygdaloid region's superior dorsal boundary are less distinct. Nonetheless, our protocols effectively traced most of the claustrum volume, though caution is advised when applying them to conventional-resolution MRIs. Third, our reliability assessment used a limited sample size, which may be insufficient for studying a structure with high morphological variability. However, bilateral claustrum DSC remained consistently high (inter-rater: 0.776–0.872; intra-rater: 0.803–0.909), demonstrating protocol robustness. Moreover, comparable studies have reliably evaluated manual segmentation with similar sample sizes (Entis et al., 2012), supporting our protocol's reliability. Fourth, our protocol was developed using only young adult samples. While substantial age-related differences in claustrum morphology are unlikely, rapid growth during late adolescence (Bennett & Baird, 2006) suggests that anatomical investigations across a wider age range may be needed to adapt the protocol for pediatric brains. The current protocol may serve as a foundation for such future studies. Finally, since the fully automatic parcellation of claustral subregions relies on FreeSurfer's segmentation, its accuracy may be compromised by FreeSurfer-generated errors. In our study, five subjects exhibited minor errors, such as mislabeling of a small number of anterior-inferior temporal claustrum voxels extending beyond the amygdala's anterior boundary (see Supplementary Figure 1). These errors were minimal and did not significantly affect reliability. However, final visual inspection and correction, as in the semi-automatic approach, are recommended to enhance accuracy.

*Conclusion*

To our knowledge, it is the first study that developed comprehensive manual segmentation protocols of the human claustrum and the automatic parcellation method for the three subregions, reporting the volume of the entire and subregional claustrum *in vivo*. Due to the thin and complicated morphology of the claustrum, approaches utilizing conventional MRIs with limited resolutions ($\geq 1.0$ mm³ voxels) have shown limited accuracy in delineating the entire claustrum. Employing the state-of-the-art cellular-resolution brain atlas and high-resolution T1-weighted MRIs with excellent gray-white matter contrast, we were able to delineate the entire claustrum, including its inferior portion known as the temporal claustrum, and achieved high inter- and intra-rater reliability using our segmentation protocols. Although it does not resolve the claustrum's finest microstructures, it enables reliable delineation of the overall anatomy of the claustrum using increasingly available submillimeter-resolution MRI. As such, it serves as a valuable resource for the neuroimaging community and supports future MRI-based studies of the claustrum. Therefore, our claustrum segmentation protocol will significantly contribute to future studies of systematic, large-scale investigations of the anatomy and the functions of the human claustrum in normal and pathological populations.

**Data Availability Statement**: The original MRI data, claustrum label files generated for this study, and the MATLAB codes for claustrum parcellation have been deposited in the Open Science Framework (OSF) repository under the DOI: 10.17605/OSF.IO/JKFQ2.